\documentclass{ws-procs9x6}

\begin{document}

\title{NONABELIAN GLOBAL CHIRAL SYMMETRY\\
REALISATION IN THE TWO-DIMENSIONAL\\
$N$ FLAVOUR MASSLESS SCHWINGER MODEL}

\author{LAURE GOUBA,$^{1,2}$ JAN GOVAERTS$^{1,3}$ and
M. NORBERT HOUNKONNOU$^{1,2}$}

\address{
$^{1}$International Chair in Mathematical Physics and Applications (ICMPA),\\
University of Abomey-Calavi,\\
072 B.P. 50, Cotonou, Republic of Benin\\
E-mail: norbert\_hounkonnou@cipma.net\\
\vspace{5pt}
$^{2}$Unit\'e de Recherche en Physique Th\'eorique (URPT),\\
Institut de Math\'ematiques et de Sciences Physiques (IMSP),\\
Universit\'e d'Abomey-Calavi,\\
01 B.P. 2628, Porto-Novo, R\'epublique du B\'enin\\
E-mail: lrgouba@yahoo.fr\\
\vspace{5pt}
$^{3}$Center for Particle Physics and Phenomenology (CP3),\\
Institute of Nuclear Physics, Catholic University of Louvain,\\
2, Chemin du Cyclotron, B-1348 Louvain-la-Neuve, Belgium\\
E-mail: Jan.Govaerts@fynu.ucl.ac.be}

\begin{abstract}
The nonabelian global chiral symmetries of the two-dimensional $N$ 
flavour massless Schwinger model are realised through 
bosonisation and a vertex operator construction.
\end{abstract}

\vspace{30pt}

\begin{center}
To appear in the Proceedings of the Fourth International Workshop on Contemporary
Problems in Mathematical Physics,\\
November 5$^{\rm th}$-11$^{\rm th}$, 2005, Cotonou (Republic of Benin),\\
eds. J. Govaerts, M. N. Hounkonnou and A. Z. Msezane\\
(World Scientific, Singapore, 2006).
\end{center}

\vspace{30pt}

\begin{flushright}
CP3-06-10\\
ICMPA-MPA/2006/27
\end{flushright}

\bodymatter

\clearpage

\section{Introduction}
\label{Sec1}

Recently,\cite{lau} the quantum two-dimensional $N$ flavour massless Schwinger 
model has been revisited without any gauge fixing but using the method of 
Dirac quantisation. The physical spectrum of this ideal
``theoretical laboratory" for nonperturbative quantum field theory is known, 
and consists of one massive pseudoscalar field which is essentially 
the electric field with squared mass $(N/\pi) e^2$, $e$ being the U(1) 
gauge coupling constant, and $(N-1)$ massless scalar fields, 
none of which are interacting. At the quantum level, this model has
SU(N)$_{-}\times$SU(N)$_{+}\times$U(1)$_V$ chiral symmetries,
where the factor SU(N)$_{-}\times$SU(N)$_{+}$ mixes separately
each the chiral Dirac fermionic field components of given chirality,
while the factor U(1)$_V$ is a common phase symmetry associated to the
total fermionic number. The associated SU(N)$_\pm$ and U(1)$_V$ 
Noether currents are bosonised. Using implicitly techniques from 
two-dimensional conformal field theory and string theory developed 
in the 1990's,\cite{God} one may construct vertex operators in direct 
relationship with these global chiral symmetries. From the modes of 
both these bosonised Noether 
currents and these vertex operators, we realise two commuting
affine Kac--Moody algebras, of which the zero modes of the vertex operators
are shown to correspond to the generators of the nonabelian global 
chiral symmetries.

This paper is organised as follows. In Sec.~\ref{Sec2}, we introduce 
the two-dimensional $N$ flavour massless Schwinger model. 
In Sec.~\ref{Sec3}, we identify the chiral symmetries of the model
and specify our notations by also defining the Hilbert space in which we work.
In Sec.~\ref{Sec4}, we construct the relevant vertex operators. 
Section~\ref{Sec5} is devoted to the Kac--Moody algebra. In 
Sec.~\ref{Sec6}, we realise the nonabelian global symmetries. 
Concluding remarks appear in Sec.~\ref{Sec7}.

\section{The Two-Dimensional $N$ Flavour Massless Schwinger Model}
\label{Sec2}

\subsection{The classical formulation}
\label{Sec2.1}

Let us consider the two-dimensional $N$ flavour massless Schwinger model 
with a dynamics described by the following Lagrangian density,
\begin{eqnarray}\nonumber
{\mathcal{L}} &=& -\frac{1}{4}F_{\mu\nu}F^{\mu\nu}
+\frac{i}{2}\sum_{j=1}^{N}\bar{\psi}^j\gamma^\mu\partial_\mu\psi^j\\
& & - \frac{i}{2}\sum_{j=1}^{N}\partial_\mu\bar\psi^j\gamma^\mu\psi^j
-e\sum_{j=1}^N\bar\psi^j\gamma^\mu A_\mu\psi^j,
\end{eqnarray}
where $F_{\mu\nu} = \partial_\mu A_\nu -\partial_\nu A_\mu$.
As usual, the spacetime coordinate indices take the values $\mu = (0,1)$,
while the Minkowski spacetime metric signature is 
${\rm diag}\,\eta_{\mu\nu} = (+,-)$. We also assume a system of units 
such that $c= 1 = \hbar$. This dynamics is singular, and, reducing 
the second-class constraints through the introduction of
Dirac brackets, the fundamental first-class Hamiltonian reads as
\begin{eqnarray}\nonumber
{\mathcal{H}} &=& \frac{1}{2}\pi_1^2 -\frac{i}{2}\sum_{j=1}^{N}\psi^{j^\dagger}
\gamma_5(\partial_1 -ie A^1)\psi^j\\
& & + \frac{i}{2}\sum_{j=1}^{N}(\partial_1 +ie A^1)
\psi^{j^\dagger}\gamma_5\psi^j -\partial_1[A^0\pi_1],
\end{eqnarray}
where the first-class constraint, related to the U(1) local gauge invariance
of the system, is simply
\begin{eqnarray}
\sigma &=& \partial_1\pi_1 + e\sum_{j=1}^{N}\psi^{j^\dagger}\psi^j.
\end{eqnarray}

\subsection{Quantum formulation}
\label{Sec2.2}

Through canonical quantisation and within the Schr\"odinger picture,
bosonisation of the Dirac fermionic operators is achieved as
\begin{eqnarray}
\hat{\psi}^j_{\pm}(z) 
= e^{i\pi\lambda\sum_{k=1}^{N}(\alpha_{\pm,k}^j\hat{p}_\pm^k
+\beta_{\pm}^j\hat{p}_\mp^k)} :e^{\pm i\lambda\hat{\phi}_\pm^j(z)}: ,
\end{eqnarray}
where the first factor on the r.h.s of this expression  represents 
the Klein factor necessary in order to have fermionic operators of different 
flavours or chiralities that anticommute with one another,
while the quantities $\hat{\phi}_\pm^j(z)$ are real chiral bosons. 
Applying the point splitting regularisation procedure and some
field redefinitions, the fundamental quantum Hamiltonian is given as
\begin{eqnarray}\nonumber
\hat{\mathcal{H}} &=& \frac{1}{2}\hat{\pi}_\varphi^2 
+\frac{1}{2}\mu^2\hat{\varphi}^2
+\frac{1}{2}(\partial_1\hat\varphi)^2 
+\frac{1}{2}\left(\frac{\hat\sigma}{\mu}\right)^2
+ \left(\frac{\hat\sigma}{\mu}\right)\partial_1\hat\varphi \\
& & + \frac{1}{4\pi}\sum_{j=1}^{N-1}\left(\partial_1\hat\Phi^j_{+}\right)^2
+ \frac{1}{4\pi}\sum_{j=1}^{N-1}\left(\partial_1\hat\Phi_{-}^j\right)^2,
\end{eqnarray}
where $\hat\varphi$ is essentially (up to a normalisaton factor)
the electric field with squared mass $\mu^2 = (N/\pi) e^2$.
Here, $\hat{\Phi}_\pm^j $ are massless chiral bosons defined by
\begin{eqnarray}
\hat\Phi_\pm^j = \frac{1}{\sqrt{j(j+1)}}
\left(\sum_{k=1}^{j}\hat{\phi}_\pm^k - j\hat{\phi}^{j+1}_\pm\right),\qquad
j\in\{1,...,(N-1)\}.
\end{eqnarray}

\section{Chiral Symmetries}
\label{Sec3}

\subsection{SU(N)$_\pm$ currents} 
\label{Sec3.1}

The SU(N)$_\pm$ currents associated to the chiral symmetries are 
defined by
\begin{eqnarray}
\hat{J}_\pm^{a\mu} =\sum_{i,j=1}^{N}\hat{\bar{\psi}}^i_\pm\gamma^\mu 
\left(\mp\lambda L\sqrt{2}\lambda^a\right)_{ij}\hat{\psi}_{\pm}^j,
\qquad \mu =0,1,
\end{eqnarray}
where the matrices $\lambda^a$, $a\in \{1,...,(N^2-1)\}$, are
$(N^2-1)$ independent hermitian traceless matrices spanning the
SU(N) algebra. These matrices are a generalisation of the Pauli matrices 
in the SU(2) case or the \mbox{Gell-Mann} matrices in the SU(3) case. 
In particular, the matrices associated to the Cartan subalgebra 
U(1)$^{N-1}$ of the Lie algebra $su(N)$ may be chosen to be given by
\begin{eqnarray}\label{19}
\left(\lambda^i\right)_{jl} =
\frac{1}{\sqrt{2i(i+1)}}\left(\sum_{k=1}^{i}\delta_{jk}\delta_{lk} 
-i\delta_{j,i+1}\delta_{l,i+1}\right),\quad\!\!\!
i\in\left\{1,...,(N-1)\right\},
\end{eqnarray} 
leading to the following associated currents,
\begin{eqnarray}
\hat{J}_\pm^{i0} &=&
(\mp\lambda L\sqrt{2})\sum_{l,j=1}^{N}{\hat{\psi}_\pm}^{j^\dagger}
\left(\lambda^i\right)_{jl}\hat{\psi}_\pm^l,\\
\hat{J}_\pm^{i1} &= &
\left(\mp\lambda L\sqrt{2}\right)\sum_{l,j =1}^{N}{\hat{\psi}_\pm}^{j^\dagger}
\gamma_5\left(\lambda^i\right)_{jl}\hat{\psi}_\pm^l.
\end{eqnarray}
Through the bosonisation procedure of the fermionic operators, one finds,
\begin{eqnarray}
\hat{J}_\pm^{i0} =\left(\pm\frac{L}{2\pi}\right)\partial_1\hat{\Phi}_\pm^i,
\qquad
\hat{J}_\pm^{i1} =\left(-\frac{L}{2\pi}\right)\partial_1\hat{\Phi}_\pm^i.
\end{eqnarray}
Let us consider the U(1)$^{N-1}$ currents given by
\begin{eqnarray}
\hat{J}_\pm^i(x) = 
\left(\pm\frac{L}{2\pi}\right)\partial_1\hat\Phi_\pm^i(x),\qquad
i\in\{1,...,(N-1)\}.
\end{eqnarray}
In terms of modes, these currents may be written as
\begin{eqnarray}\label{n2}
\hat{J}_\pm^i(z) = \hat{P}_\pm^i +\sum_{n\ge1}
\left({\hat{J}_{\pm,n}}^{i^\dagger}z^n +
\hat{J}_{\pm,n}^i z^{-n}\right),
\end{eqnarray}
where the modes are function of the modes of the $(N-1)$ real bosonic 
fields $\Phi_\pm^i$, and satisfy the following algebra
\begin{eqnarray}\label{36}
\left[\hat{J}_{\pm,n}^i\:,\:\hat{J}_{\pm,m}^{j^\dagger}\right] 
=n\delta_{ij}\delta_{mn}.
\end{eqnarray}

\subsection{The quantum Hilbert space}
\label{Sec3.2}

{}From the algebra (\ref{36}), we conclude that the operators
$\hat{J}_{\pm,n}^i$ and $\hat{J}_{\pm,n}^{j^\dagger}$, 
with $i,j = \{1,\cdots,(N-1)\}$ and $n = 1,\cdots,\infty$,
form a set of independent harmonic oscillators. Therefore, the state space 
considered here is a Fock space built up from the simultaneously normalized 
vacua of all these oscillators, $|0\rangle$ ,
\begin{eqnarray}
\hat{J}_{\pm,n}^j |0\rangle =0,\quad n>0,\quad
\hat{P}_\pm^j |0\rangle =0.
\end{eqnarray}
Let us consider $\Sigma$, the set of roots of the Lie algebra $su(N)$,
and $\Lambda_\Sigma$ the root lattice of $su(N)$. States 
$\vert \lambda \rangle$ can be added to the above states by acting with 
the plane wave operator $\displaystyle{e^{i\lambda\cdot\hat{Q}_\pm}}$. We denote
\begin{eqnarray}\label{70}
|\lambda\rangle = e^{i\lambda\cdot\hat{Q}_\pm}\vert 0\rangle,
\end{eqnarray}
where $\lambda \in \Lambda_\Sigma$. Later we shall assume that all
$\alpha\in \Sigma$ have length $\sqrt{2}$.

\section{Vertex Operators}
\label{Sec4}

Given any complex number $z$ and any root $\alpha$, 
let us introduce the vertex operator\cite{God} 
\begin{eqnarray}\label{33}
\hat{U}_\pm^\alpha(z) =
z^{\frac{\alpha^2}{2}} :e^{i\alpha\cdot\hat{\Phi}_\pm(z)}:
\end{eqnarray}
where
\begin{eqnarray}
\alpha\cdot\hat\Phi_\pm(z) = \sum_{j=1}^{N-1}\alpha^j\hat\Phi_\pm^j(z).
\end{eqnarray}
In order to make (\ref{33}) single valued in $z$, the following condition 
is required,
\begin{eqnarray}
\left( \frac{\alpha^2}{2} +\alpha\cdot\hat{P}_\pm\right)\in \mathbb{Z}.
\end{eqnarray}
Therefore (\ref{33}) is analytic and has a Laurent expansion 
\begin{eqnarray}
\hat{U}_\pm^\alpha(z) =\sum_{m\in\mathbb{Z}}\hat{U}_{\pm,m}^\alpha z^{-m},
\end{eqnarray}
where the modes can be written as
\begin{eqnarray}
\hat{U}_{\pm,m}^\alpha = \oint_0\frac{dz}{2i\pi z}z^m\hat{U}_\pm^\alpha (z)
\end{eqnarray}
and satisfy the hermiticity condition
\begin{eqnarray}
{\hat{U}_{\pm,m}}^{\alpha^\dagger} =\hat{U}_{\pm,-m}^{-\alpha}.
\end{eqnarray}

\section{The Kac--Moody Algebra}
\label{Sec5}

\subsection{Almost commutation relations}
\label{Sec5.1}

Using the modes of currents and those of vertex operators, we
can build almost commutation relations by
\begin{eqnarray}\label{59}
\left[\hat{J}_{\pm,n}^i\:,\:{\hat{J}_{\pm,m}}^{j^\dagger}\right] 
=  m\delta_{ij}\delta_{m,n},\quad
\left[\hat{J}_{\pm,n}^i\:,\:\hat{U}_{\pm,m}^\alpha\right] 
= \alpha^i\hat{U}_{\pm,m+n}^\alpha ,
\end{eqnarray}
\begin{eqnarray}\label{61}
\hat{U}_{\pm,m}^\alpha\hat{U}_{\pm,n}^\beta 
-(-1)^{\alpha\cdot\beta}\hat{U}_{\pm,n}^\beta\hat{U}_{\pm,m}^\alpha 
= \left\{\begin{array}{lll}
0&\ {\rm if}\ \alpha\cdot\beta\ge 0,\\
\hat{U}_{\pm,m+n}^{\alpha+\beta} &\ {\rm if}\ \alpha\cdot\beta =-1,\\
\alpha\cdot\hat{J}_{\pm,m+n}   & \\
+m\delta_{m+n,0} &\ {\rm if}\ \alpha\cdot\beta =-2,
\end{array}\right.
\end{eqnarray}
where $\displaystyle{\alpha\cdot\hat{J}_{\pm,m+n} 
=\sum_{j=1}^{N-1}\alpha^j\hat{J}^j_{\pm,m+n}}$.

In order to have commutation relations, we have
to correct the sign $(-1)^{\alpha\cdot\beta}$ which appears in (\ref{61}).

\subsection{Sign compensation}
\label{Sec5.2}

Let us set $V=\left({\mathcal{F}},\: 
\left\{e^{i\lambda\cdot\hat{Q}_\pm}\right\}_{\lambda\in\Lambda_\Sigma}\right)$,
where ${\mathcal{F}}$ denotes the Fock space. We define a sign compensation
operator\cite{Fren} by
\begin{eqnarray}\label{74}
C_{\pm,\alpha} =
\sum_{\beta\in\Lambda_\Sigma}\epsilon(\alpha,\beta)
\vert\beta\rangle\langle\beta\vert,
\end{eqnarray}
which only acts on the wave plane factor and satisfies the following conditions 
\begin{eqnarray}\label{71}
\epsilon(\alpha,\beta) \in \{-1,1\}&,&\quad
\epsilon(\alpha,\beta) 
= (-1)^{\alpha\cdot\beta +\alpha^2\beta^2}\epsilon(\beta,\alpha),\\ \label{73}
\epsilon(\alpha,\beta)\epsilon(\alpha+\beta,\gamma) &=& 
\epsilon(\alpha,\beta+\gamma)\epsilon(\beta,\gamma).
\end{eqnarray}
Let us set
\begin{eqnarray}
\hat{E}_{\pm,n}^\alpha = \hat{U}_{\pm,n}^\alpha C_{\pm,\alpha},\qquad
\hat{E}_{\pm,m}^\beta = \hat{U}_{\pm,m}^\beta C_{\pm,\beta}.
\end{eqnarray}
We then have the following commutation relations, known as
the affine Kac--Moody algebra (in fact, we obtain two such
algebras, one for each of the chiral sectors of the model, which
commute with one another),
\begin{eqnarray}
\left[\hat{J}_{\pm,m}^i,\:{\hat{J}_{\pm,n}}^{j^\dagger}\right] 
= m\delta_{ij}\delta_{m,n}, \qquad
\left[\hat{J}_{\pm,m}^i,\:\hat{E}_{\pm,n}^\alpha\right] =
\alpha^i\hat{E}^\alpha_{\pm,m+n},
\end{eqnarray}
\begin{eqnarray}
\left[\hat{E}_{\pm,m}^\alpha,\:\hat{E}_{\pm,n}^\beta\right]
=\left\{ \begin{array}{lll}
\epsilon(\alpha,\beta) \hat{E}_{\pm,m+n}^{\alpha +\beta}
&\ {\rm if}\ \alpha\cdot\beta =-1,\\
\alpha\cdot\hat{J}_{\pm,m+n} + m\delta_{m,n} 
&\ {\rm if}\ \alpha\cdot\beta =-2,\\
0 &\ {\rm if}\ \alpha\cdot\beta \ge 0.
\end{array}\right.
\end{eqnarray}

\section{Nonabelian Global Chiral Symmetries}
\label{Sec6}

We are now ready to realise the nonabelian global chiral symmetries of
the model. In fact, having defined the affine Kac--Moody algebra, 
it is well known from the literature\cite{God} that the following algebra 
is isomorphic to the Lie algebra $su(N)$,
\begin{eqnarray}
\hat{J}_{\pm,0}^i,\qquad \hat{E}^\alpha_{\pm,0},
\qquad 1\le i\le (N-1),\quad \alpha\in\Lambda. 
\end{eqnarray}
The Cartan subalgebra is generated 
by $\hat{J}_{\pm,0}^i$, $i\in\{1,...,(N-1)\}$,
while the nonabelian global symmetries are realised by
$\hat{E}_{\pm,0}^\alpha$, $\alpha \in \Lambda$. 

\section{Concluding remarks}
\label{Sec7}

We have realised the nonabelian global chiral symmetries of the
two-dimensional $N$ flavour massless Schwinger model through
bosonisation of the massless Dirac spinors. Our results generalise 
the recent work of Michael Creutz.\cite{Mich} An important aspect 
relative to the nature of the action of these nonabelian global 
chiral symmetries on the one-particle quantum states, and
beyond, of the model, would merit further investigations.

\section*{Acknowledgements}

The authors acknowledge the Agence Universitaire de la Francophonie (AUF)
and the Belgian Cooperation CUD-CIUF/UAC for financial support. L. G. is
presently supported through a Ph.D. Fellowship of the Third World Organisation
for Women in Science (TWOWS, Third World Academy of Science). 
J. G. acknowledges the Abdus Salam International Centre for Theoretical
Physics (ICTP, Trieste, Italy) Visiting Scholar Programme 
in support of a Visiting Professorship at the International Chair 
in Mathematical Physics and Applications (ICMPA). The work of J. G. 
is partially supported by the Belgian Federal Office for Scientific, 
Technical and Cultural Affairs through the Interuniversity Attraction 
Pole (IAP) P5/27.

\end{document}